\documentstyle[aps,twocolumn,epsfig]{revtex}
\begin{document}
\draft
\title{Single  photons from Pb+Pb collisions at CERN SPS, QGP vs.
hadronic gas}

\author{\bf A. K. Chaudhuri\cite{byline}}
\address{ Variable Energy Cyclotron Centre\\
1/AF,Bidhan Nagar, Calcutta - 700 064\\}

\maketitle
\begin{abstract}
In  a hydrodynamic model, we have analyzed the direct photon data
obtained by the WA98 collaboration in 158 A GeV Pb+Pb  collisions
at  CERN  SPS.  The  transverse expansion of the system was taken
into account. Two scenarios, (i) formation of  quark-gluon  plasma
and  (ii)  formation of hot hadronic gas, were considered. Both the
scenarios describe the data equally well.
 However,  hadronic  gas  scenario  require  very   high   initial
temperature  ($\sim$  300  MeV)  and  it is difficult to conceive
existence of hadron gas at that high temperature. If the hadronic
fluid has small radial velocity (0.2c-0.3c) initially,  the  data
are  well  explained in  the hadronic gas scenario with reasonable
initial temperatures. \end{abstract}


Recently  WA98  collaboration  has  published their single photon
emission data  for  158  A  GeV  Pb+Pb  collisions  at  CERN  SPS
\cite{wa98}.  Much  interest was aroused after the publication of
the WA80 preliminary results  of  the  S+Au  single  photon  data
\cite{sa93},  as  it  was hoped that the they can be a conclusive
probe of the much debated quark-gluon plasma (QGP), expected to be
produced in relativistic heavy ion  collisions.  The  preliminary
data   were  analyzed  by  several  authors.  Xiong  and  Shuryak
\cite{sh94} analyzed the data assuming a  mixed  phase  formation
and  found  excess  photons.  Srivastava  and  Sinha  \cite{sr94}
analyzed the data considering two possible  scenarios  after  the
collision,  one  with  the  phase  transition  to  QGP, the other
without it. It was claimed that the data were explained  only  in
the   phase   transition  scenario.  We  had  also  analyzed  the
preliminary versions of the WA80 direct photon  data  \cite{ch95}.
It  was  shown  that formation of {\em viscous} hadron gas in the
initial state, can explain the data. The revised version  of  the
data  \cite{al96} were also analyzed by several authors including
us\cite{ch00}. It was concluded that the data are  not  sensitive
enough  to  discriminate between the two alternate pictures, e.g.
formation of quark-gluon plasma and  formation  of  hot  hadronic
gas.

Several  authors have analyzed the recent WA98 single photon data
\cite{sr00,ja00,ga00,pe00}. Peressounko et al \cite{pe00} 
 concluded that
the data could be explained only with a small initial radial
velocity. Also the data could not distinguish between a QGP
and a hot hadronic gas scenario.
Srivastava et al \cite{sr00} 
could fit the data in the QGP scenari,
without initial radial velocity. But the two loop photon rate used
by them was not corrected for a factor of 4.  They  did  not  considered  the
hadronic   scenario.  They  argued  that  in  the  pure  hadronic
scenario, initial temperature  of  the  hadronic  fluid  will  be
large.  Hadronic  density will be $\sim$ 10 hadrons/$fm^3$. It is
unphysical to consider hadronic  gas  at  such  a  high  density.
Initial  Hadronic  scenario was considered in \cite{ja00}. It was
concluded that  only  hadronic  model  with  medium  modification
(hadrons were formed with zero mass) could explain the data.

In  the  present  paper we analyze the WA98 single photon data in
the no phase transition (NPT) scenario. A  hot  hadronic  gas  is
assumed  to be formed in the initial state. It expands, cools and
freezes out at freeze-out temperature ($T_F$). It will  be  shown
that  WA98  single  photon  data  could be well explained in this
scenario, with reasonable hadron density, if one assumes a  small
initial   fluid   velocity.   In   \cite{pe00}  initial  velocity
distribution was assumed to be  linearly  increasing.  We  assume
that  at  the  beginning of one fluid hydrodynamic stage, initial
radial  velocity  follows  the  distribution  of  initial  energy
density,  which  we assume to be of Woods-Saxon form, with radius
and diffuseness  parameters,  $R=6.4  fm$  and  $a=0.54  fm$.  In
\cite{pe00}  hadronic equation of state comprises all the hadrons
in the particle data book. We choose to use hadrons up to  mass  2
GeV.  The  cut off is arbitrary. The resulting equation of sate is
reasonably well described by $p_h  =  a_h  T^4$,  with  $a_h=59.5
\pi^2/90$.  Fig.1 compares the analytic expression with numerical
results. Analytic expression overestimate the pressure in the temperature
region of 200-400 MeV. However, we have used the analytic expression
only to obtained the initial temperature of hadron gas, and the
error introduced due to the approximation is around 10-15\%.
To be complete, we also analyze the data  in  the  phase
transition  (PT)  scenario,  when  QGP  is  formed in the initial
state. The equation of state for QGP was assumed to  be  $p_q=a_q
T^4  -B$  with  $a_q=42.25  \pi^2/90$.  The  bag constant $B$ was
obtained from the Gibbs condition $p_{QGP} (T_c)= p_{had}(T_c)$.

We  solve the hydrodynamic equations $\partial_\mu T^{\mu \nu}=0$
in   3+1   dimension   assuming    cylindrical    symmetry    and
boost-invariance  in  the  longitudinal  direction.  The relevant
equations are well known \cite{ge86,ja93} and are not  reproduced
here.  The  inputs  of  the hydrodynamic equations are the initial
energy  density  or  temperature  ($T_i$)  and  radial   velocity
($v_r^{ini}$)   at   (proper)  time  $\tau_i$.  $\tau_i$  is  the
thermalisation time beyond which hydrodynamics became applicable.
For a given $\tau_i$ the initial temperature $T_i$ of  the  fluid
(hadronic  gas  or  QGP)  can be obtained by relating the entropy
density  with  the  observed  pion  multiplicity  (assuming  pion
decoupling to be adiabatic) \cite{hw85},

\begin{equation} \label{1}
T^3_i\tau_i=\frac{1}{\pi R^2_A} \frac{c}{4a_{q,h}}\frac{dN}{dy}
(b=0)
\end{equation}

\noindent where $c=2\pi^4/45\zeta(3)$ and $R_A$ is the transverse
radius of the system (assumed to be 6.4 fm for Pb+Pb collisions).
$b=0$  corresponds  to  central  collisions. In table 1., we have
shown the initial temperatures as obtained from eq.\ref{1} in the
two considered scenarios, the  QGP  and  the  hot  hadronic  gas.
$dn/dY$  was  assumed  to  be 750 \cite{sr00}. It can be seen, in
both  the  scenarios,  initial   temperatures   are   comparable.
Corresponding hadron densities  are  also  shown  in table 1. For
$\tau_i$=0.2 and 0.4 fm, it is $\sim$  10  and  2.9  hadrons  per
$fm^{-3}$  respectively. It is unlikely than at such high density
hadrons can retain their identity. For larger  $\tau_i$,  initial
temperatures  are  comparatively  small  and  the  densities have
acceptable values.  The  other  parameter  for  the  hydrodynamic
evolution   with  transverse  expansion  is  the  initial  radial
velocity $v_r^{ini}$. It is customary to assume that the  initial
$v_r^{ini}$=0.  However,  it  is  possible that the fluid (QGP or
hadronic gas) possess some small radial velocity at initial  time
$\tau_i$.  As  will  be shown here, initial small radial velocity
can affect the photon spectra considerably.

For the single photons from hadronic gas we include the following
processes,

(a) $\pi\pi \rightarrow \rho \gamma$,
(b)  $\pi  \rho \rightarrow \pi \gamma$,
(c) $\omega \rightarrow \pi \gamma$,
(d) $\rho \rightarrow \pi \pi \gamma$
(e) $\pi \rho \rightarrow A_1 \rightarrow \pi \gamma$

\noindent rates for which are well known \cite{na92,xi92}.

Rate  of  production  of  hard photons from QGP were evaluated by
Kapusta et al \cite{ka91}. To one loop order,

\begin{equation}\label{2}
E  \frac{dR}{d^3p}  =
\frac{1}{2\pi^2} \alpha \alpha_s \sum_f e^2_f T^2 e^{-E/T}
\ln(\frac{cE}{\alpha_s T})
\end{equation}

\noindent  where  the  constant  $c\sim 0.23$. The summation runs
over the flavors of the quarks and $e_f$ is the electric  charge
of the quarks in units of charge of the electron.

Recently  Aurenche  et  al\cite{au98} evaluated the production of
photons in a QGP.  At  two  loops  level  Bremsstrahlung  photons
$(qq(g)  \rightarrow  qq(g)  \gamma$  found  to be dominating the
compton and annihilation  photons.  However,  their  calculations
overestimated    the    photon   yield   by   a   factor   of   4
\cite{au98a,st01}.

The   rate   of  production  of  photons  due  to  Bremsstrahlung
(corrected for the factor of 4) was evaluated by them as,

\begin{equation}\label{3}
E  \frac{dR}{d^3p} =\frac{1}{4} \frac{8}{\pi^5} \alpha \alpha_s \sum_f e^2_f
\frac{T^4}{E^2} e^{-E/T} (J_T - J_L) I(E,T)
\end{equation}

\noindent  where  $J_T  \sim  4.45$  and $J_L \sim -4.26$ for two
flavors and 3 colors of quarks. For 3 flavor quarks, $J_T  \sim
4.8$ and $J_L \sim -4.52$. $I(E,T)$ stands for,

\begin{eqnarray}\label{4}
I(E,T)     =     &&[3\zeta(3)     +    \frac{\pi^2}{6}\frac{E}{T}
+(\frac{E}{T})^2 \ln 2  +4  Li_3(-e^{-|E|/T})  \nonumber  \\  &&+
2Li_2(-e^{-|E|/T}) -(E/T)^2 \ln(1+e^{-|E|/T})]
\end{eqnarray}

\noindent and the poly-logarith functions $Li$ are given by,

\begin{equation}\label{5}
Li_a(z) = \sum_{n=1}^{\infty} \frac{z^n}{n^a}
\end{equation}

Aurenche  et  al  \cite{au98} also calculated the contribution of
the $q\bar{q}$ with scattering, which was also  overestimated  by
the same factor of 4. The corrected rate is,

\begin{equation}\label{6}
E \frac{dR}{d^3p} =\frac{1}{4} \frac{8}{3\pi^5} \alpha \alpha_s \sum_f e^2_f
E T e^{-E/T} (J_T - J_L)
\end{equation}

We  would  like  to mention that two loop photon rate from QGP is
not  complete.  Higher  loops  contribute  to  the   same   order
\cite{au00a}.    Also   Landau-Migdal-Pomeranchuk   effect   been
neglected   \cite{au00b}.

Before  we  present  our  results  we  would like to make a brief
comment on  the  direct  QCD  photons.  Direct  QCD  photons  are
produced  from the early hard collisions of partons in the nuclei
and in Pb+Pb collisions make significant contribution to the high
$p_T$ yield. Gallmeister et al \cite{ga00}  claimed  that  prompt
photons  are  able  to  explain  the  high  $p_T$  data  in Pb+Pb
collisions. Dumitru et al \cite{du01} also arrived at  a  similar
conclusion including the nuclear broadening effects. However, this
point  is  still  controversial  due  to  uncertainties in prompt
photon emission at AA collisions. Thus Alam et al \cite{ja00} and
also Srivastava  and  Sinha  \cite{sr00}  calculated  the  prompt
photon  emission for Pb+Pb collisions. It was seen that for $p_T$
$>$2 GeV, direct QCD photons alone can describe the data within a
factor 3-8 only.

We  first  present  the  photon  spectra  obtained  in  the phase
transition scenario. As in ref.\cite{sr00} we assume the critical
temperature  to  be  $T_c$=180  MeV.  Data  were  found   to   be
insensitive  to the exact value of $T_c$ \cite{sr00}. The initial
radial velocity ($v_r^{ini}$) was assumed to be zero.  In  fig.2,
computed    photon    spectra   for   different   initial   times
$\tau_i$=.2,.4,.6 and .8 fm, with one+two loop  order  rates  are
shown.   In the inset of Fig.1, we have compared the photon yield
obtained with one loop rate and
one+two loop rate. For initial time $\tau_i$ = 0.2  fm,  two
loop  contribution  is  more than 50\% at large $p_T$. Higher
loop contributing to the  same order, photon
rates,   are     not   complete.  But  for
  $\tau_i> .2 fm$,  two  loop  contribution  is quite less and the
rates can be regarded more or less complete. We  find  that  with
corrected  photon rate, QGP scenario, unlike in ref. \cite{sr00},
do not describe the data well.  In  \cite{sr00},
photon rates were not corrected for the factor of 4. With correct
photon rate, yield is reduced by  a  factor  of  2,  compared  to
uncorrected rate.
Consequently,   while   Srivastava   et   el
\cite{sr00}  found  very  good  fit  to data for $\tau_i$-0.2 fm,
presently data are underestimated by the similar factor of 2. For
higher thermalisation times, the data are further  underestimated
particularly at high $p_T$ side. The initial temperatures are low
enough  to  produce  requisite  number  of high $p_T$ photons. In
fig.2, the dotted line  shows  the  contribution  of  direct  QCD
photons,  as  calculated  by  Alam  et al \cite{ja00}. For
  $\tau_i > 0.2  fm$,  the  hard  QCD  photons
contribute  more     than the thermal photons.
It is evident  that  for  thermalisation  time  $\tau_i$=0.2  fm,
thermal  photons together with hard QCD photons describe the data
satisfactorily. For higher thermalisation times, thermal  photons
together  with  hard  QCD photons underpredict the data. We do not
elaborate on this scenario. Just a few comments are in order. The
equation of state of the hadronic sector now consists of  hadrons
with  mass less than 2 GeV, while in ref.\cite{sr00} hadrons with
mass less than 2.5 GeV were included. The other difference is the
initial energy density profile. Srivastava and Sinha assumed  the
initial  energy  density  profile  to  follow the wounded-nucleon
distribution,  while  we  have  used  the  standard   Woods-Saxon
profile.  Despite these differences, our results are very similar
to the results obtained in \cite{sr00}. The  data  are  not
sensitive enough to the details of the calculations. The data can
not  distinguish  whether  hadronic  sector comprised with hadron
with mass less than 2.5 GeV or with mass less than 2  GeV.  Also,
the data are insensitive to the details of initial energy density
profile  so  long  they are not very different. May be at RHIC or
LHC energy, data will be sensitive on these details.

The  results  obtained  in the no phase transition scenario, when
hadronic gas is assumed to be formed in  the  initial  state  are
presented   in  fig.3.  We  have  presented  photon  spectra  for
different initial times, $\tau_i$=.2,.4,.6,.8 fm. The results are
similar to  those  obtained  in  the  QGP  scenario.  It  is  not
surprising.  As  told  earlier,  with resonance hadronic gas, the
degrees of freedom  are  comparable  to  the  QGP.  Thus  initial
temperatures  are  nearly same either in hadron gas or in QGP. It
was noted quite early that hard photon production rate are nearly
same in hadronic gas or in QGP \cite{ka91}.  Here  also  we  have
shown  the  contribution of hard QCD photons (dotted line). As in
the scenario with QGP formation, here also for $\tau_i > 0.2  fm$
hard  QCD photons contribute more to high $p_T$ photon yield than
the thermal photons. For $\tau_i$=0.2 fm, hadronic  gas  scenario
describe the data very well. For higher thermalisation times, the
description  gets poorer especially in the high $p_T$ sector. Hard
QCD photons fill up substantially in the high $p_T$  sector,  but
as stated earlier they are less by a factor of 3-8. It would seem
that  the WA98 single photon data equally well described in PT as
well as NPT scenario. However, existence of  hadronic  gas  at  a
temperature   of   300   MeV   is  extremely  unlikely.  At  this
temperature, hadron density is very large $\sim 10  fm^{-3}$.  It
is  unlikely that at such a high density hadrons can retain their
identity. Hadronic  gas  scenario  at  such  a  high  density  is
unphysical.  For higher thermalisation time $\tau_i$=0.6 fm, when
the initial temperature and density are 211 MeV and  1.3  hadrons
per  $fm^{-3}$,  though  the  scenario is physical, the data gets
underpredicted by a factor of 3-8 (taking into  consideration  of
hard  QCD  photons).  One  of  the deficiency of the model is the
neglect of  pre-equilibrium  photons,  emitted  before  $\tau_i$.
Traxler  and Thoma \cite{tr96} calculated pre-equilibrium photons
and found them order  of  magnitude  less  than  the  equilibrium
photons.    Roy   et   al   \cite{ro97}   also   calculated   the
pre-equilibrium photons. They  used  Fokker-Plank  equations  and
found  that  pre-equilibrium photons are less or at best equal to
the equilibrium photons. The  results  refers  to  RHIC  and  LHC
energies.   At  SPS  one  expects  still  lower  contribution  of
pre-equilibrium photons. At SPS, the system is far from  chemical
equilibrium  and  number  of quarks and anti-quarks will be less.
Even if pre-equilibrium photons contribute as much as equilibrium
photons,  the  WA98   data   will   remain   underpredicted   for
thermalisation time $\tau_i \sim$ 0.6 fm or more.

It  would  seem  that  though in NPT scenario good description of
data is obtained, physical consideration (i.e. very  high  hadron
density)   will   render   that   picture  unacceptable.  In  the
calculation presented  till  now,  the  initial  radial  velocity
($v_r^{ini}$)  of the fluid was assumed to be zero. However it is
possible that at initial time $\tau_i$ the fluid has  some  small
radial  velocity  $v_r^{ini}$.
Several  authors
\cite{ja00,ga00,pe00}  have  used  initial  velocity  to  fit the
direct photon data. Source of $v_r^{ini}$ may be  the  collisions
among  the  constituents,  which  lead  to the local equilibrium.
Peressounko et al \cite{pe00} used the high $P_T$ component above
2  GeV in the $\pi^0$ spectrum to argue for it. However, argument
in favor of initial fluid velocity is weak.
High $p_T$ part of the spectrum might have
come   from   hard   processes \cite{wa99}.   Also   several   hydrodynamical
calculations,  with out initial velocity, could explain a host of
experimental data \cite{hu99,ko99,ko00}. Apparently  large  $p_T$
data require a initial fluid velocity.

 In  fig.4,  we have shown the photon spectra in the
hadronic   gas   scenario    with    initial    fluid    velocity
$v_r^{ini}$=0,.1,.2,.3 (in units of c) for initial temperature of
$T_i$=211  MeV  corresponding  to  initial  time $\tau_i$=0.6 fm.
$v_r^{ini}$  had  considerable  effect  on  photon  spectra.   It
enhances the $p_T$. Good fit to data is obtained for $v_r^{ini}$
=0.3c. It is also obvious that with $v_r^{ini}$ in the ranges  of
$0.2-0.3c$,  it  will  be  possible  to  fit  the  WA98 data in a
hadronic gas scenario with physically acceptable initial time and
temperature.

Good  fit  to the data with small initial velocity bring back the
hadronic gas scenario into contention. It is no  longer  possible
to  say  that the WA98 data indicate quark-gluon plasma formation
only.

Initial fluid velocity will also affect the photon spectra in the
phase  transition  scenario. As shown here in this scenario, WA98
data are underpredicted with initial time $\tau_i >$ 0.2  fm.  It
will  be possible to fit the data with $\tau_i >$0.2 fm, if small
initial velocity is assumed. In ref. \cite{ja00} goof fit to data
was obtained for $\tau_i$=1 fm and initial velocity in the ranges
0.2-0.3c.

 To summarize, we have analyzed the recent WA98 single photon data
using  a  hydrodynamic  model. Two scenarios were considered, the
phase transition scenario where a QGP is formed  in  the  initial
state,  and  the  no phase transition scenario where hot hadronic
gas is formed initially. Both the scenarios gave good description
to the data. QGP scenario  require  that  the  initial  time  and
temperature  of  the QGP is 0.2 fm and 340 MeV respectively. The
hadronic gas scenario also require an initial time of 0.2 fm  and
temperature  of  304  MeV. As the hadron density is very large at
this temperature, it would seem that the data is described by QGP
only. However, it  was  shown  that  with  small  initial  radial
velocity  in  the  range  0.2c-0.3c,  the  WA98  data can be well
described  in  the  hadronic  scenario  with  reasonable  initial
temperature.  The  present  analysis  thus suggests that the WA98
single photon data are not conclusive. It  can  not  discriminate
between two alternate scenarios currently in vogue.

\begin{table}
\caption{The  initial temperature of the QGP and the hot hadronic
gas for different initial times  $\tau_i$.  Also  shown  are  the
corresponding hadron density }
\label{table1}
\begin{tabular}{cccc}
$\tau_i(fm)$   &   $T^{QGP}_i   (MeV)$   &  $T^{Had}_i  (MeV)$  &
$\rho^{Had}_i (fm^{-3})$\\
\tableline
0.2 & 341 & 304 &10.41\\
0.4 & 271 & 242 &2.89\\
0.6 & 237 & 211&1.32\\
0.8 & 215 & 192&0.78\\
1.0 & 200 & 178&0.52\\
\tableline
\end{tabular}
\end{table}
\eject
\begin{figure}
\centerline{\psfig{figure=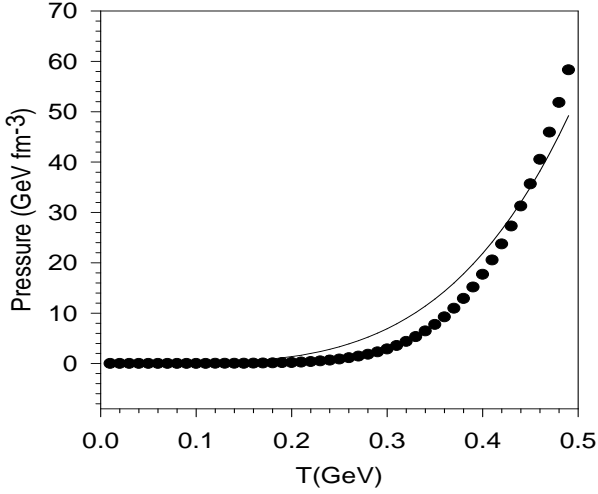,height=12cm,width=10cm}}
\vspace{-4cm}
\caption{Pressure  as  a function of temperature for the hadronic
gas comprising hadrons with mass less than 2 GeV. The solid  line
is a fit using $a_h T^4$, $a_h=59.5$.}
\end{figure}
\begin{figure}
\centerline{\psfig{figure=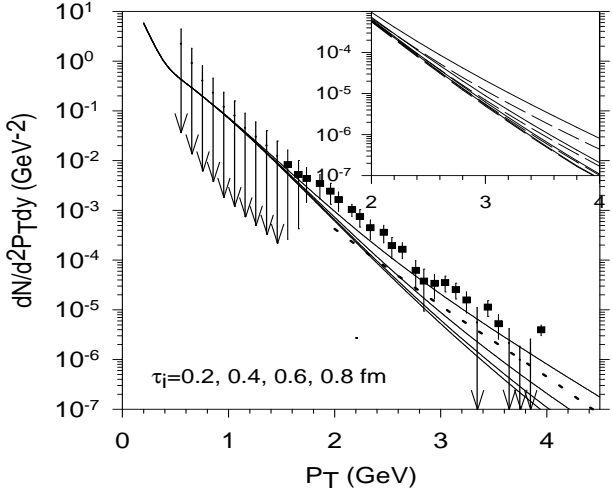,height=12cm,width=10cm}}
\vspace{-3cm}
\caption{The single photon yield in the phase transition scenario
for  four  different  initial  times,  $\tau_i$'s, .2,.4,.6,.8 fm
(from top to bottom). Corresponding temperatures  are  listed  in
table  1.  Experimental points are also shown. The dotted line is
the direct QCD photons calculated in ref.\protect\cite{ja00}.
In the inset, we have compared the photon yield obtained with one
loop rate (dashed lines) with the one+two loop rate.}
\end{figure}
\begin{figure}
\centerline{\psfig{figure=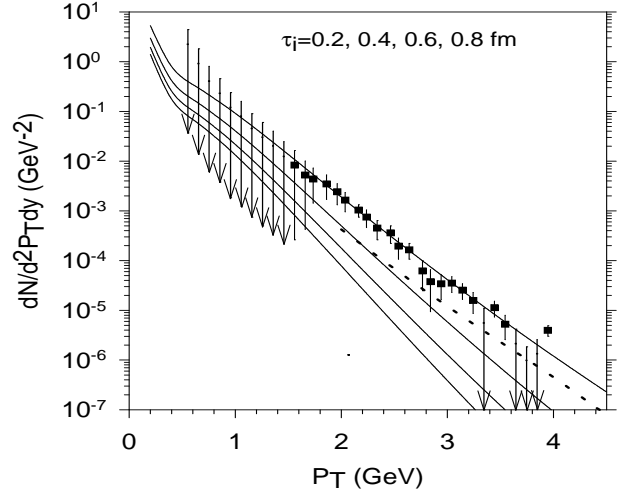,height=12cm,width=10cm}}
\vspace{-4cm}
\caption{The  single  photon  yield  in  the  no phase transition
scenario for four different $\tau_i$'s, .2,.4,.6,.8 fm (from  top
to  bottom).  Corresponding  tempertures  are  listed in table 1.
Experimental points are also shown. The dotted line is the direct
QCD photons calculated in ref.\protect\cite{ja00}}
\end{figure}
\begin{figure}
\centerline{\psfig{figure=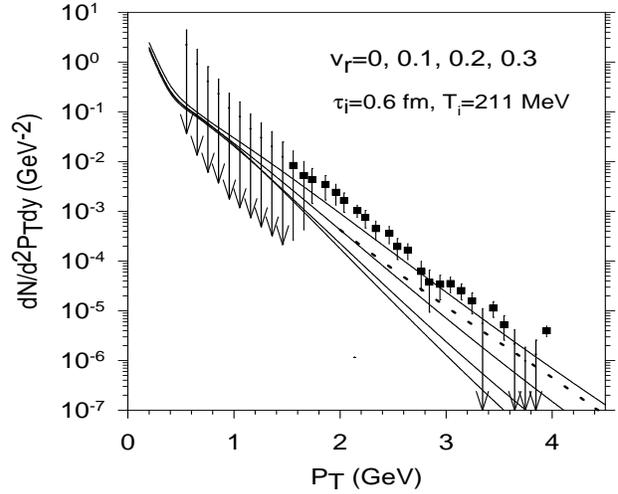,height=12cm,width=10cm}}
\vspace{-4cm}
\caption{The  single  photon  yield  in  the  no phase transition
scenario     for     four      different      initial      radial
velocity$v_r^{ini}$=0,.1,.2,3  (in  units of c). The initial time
and  temperatures  are   $\tau_i$=.6   fm   and   $T_i$=211   MeV
respectively.   The   dotted  line  is  the  direct  QCD  photons
calculated in ref.\protect\cite{ja00}.}
\end{figure}
\end{document}